\documentclass[superscriptaddress,twocolumn,floatfix,amsmath,amssymb,pra]{revtex4}

\usepackage{amsmath,amssymb}
\usepackage{graphics}
\usepackage{epsfig}

\usepackage{MnSymbol}
\usepackage{mathrsfs}
\usepackage[usenames]{color}

\begin{document}

%\title{Molecular Toolkit for Fundamental Physics and Interferometry in Microgravity}
\title{Perspectives and Opportunities: A Molecular Toolkit for Fundamental Physics and 
Matter Wave Interferometry in Microgravity}

\author{Jos\'e P. D'Incao}
\affiliation{JILA, University of Colorado and NIST, Boulder, Colorado 80309-0440, USA}
\affiliation{Department of Physics, University of Colorado, Boulder, Colorado 80309-0440, USA}
\author{Jason R. Williams}
\affiliation{Jet Propulsion Laboratory, California Institute of Technology, Pasadena, 91109 CA, USA\footnote{ A portion of this research was carried out at the Jet Propulsion Laboratory, California Institute of Technology, under a contract with the National Aeronautics and Space Administration.}}
\author{Naceur Gaaloul}
\affiliation{Institut f\"ur Quantenoptik, Leibniz Universit\"at Hannover, Welfengarten 1, D-30167 Hannover, Germany}
\author{Maxim A. Efremov}
\affiliation{Institut f\"ur Quantenphysik and Center for Integrated Quantum Science and Technology(IQST),
Universit\"at Ulm, Albert-Einstein-Allee 11, 89081 Ulm, Germany} 
\affiliation{Institute of Quantum Technologies, German Aerospace Center(DLR), Wilhelm-Runge-Stra{\ss}e 10, 89081 Ulm, Germany}
\author{Stefan Nimmrichter}
\affiliation{Naturwissenschaftlich-Technische Fakult{\"a}t, Universit{\"a}t Siegen, 57068 Siegen, Germany}
\author{Bj\"orn Schrinski}
\affiliation{Center for Hybrid Quantum Networks (Hy-Q), Niels Bohr Institute, University of Copenhagen, Blegdamsvej 17, DK-2100 Copenhagen, Denmark}
\author{Ethan Elliott}
\affiliation{Jet Propulsion Laboratory, California Institute of Technology, Pasadena, 91109 CA, USA\footnote{ A portion of this research was carried out at the Jet Propulsion Laboratory, California Institute of Technology, under a contract with the National Aeronautics and Space Administration.}}
\author{Wolfgang Ketterle}
\affiliation{Harvard-MIT Center for Ultracold Atoms, Cambridge, Massachusetts 02138, USA}
\affiliation{Department of Physics, Massachusetts Institute of Technology, Cambridge, Massachusetts 02139, USA} 

%Confirmed co-authors = Naceur Gaaloul, Jason Williams, Jose D'Incao, Maxim (reviewing it), Bjorn and Stefan. 

\begin{abstract}
The study of molecular physics using ultracold gases has provided a unique probe into the fundamental properties of nature and offers new tools for quantum technologies. In this article we outline how the use of a space environment to study ultracold molecular physics opens opportunities for 1) exploring ultra-low energy regimes of molecular physics with high efficiency, 2) providing a toolbox of capabilities for fundamental physics, and 3) enabling new classes of matter-wave interferometers with applications in precision measurement for fundamental and many-body physics.
\end{abstract}
 
\maketitle

\section{Introduction} 

With the advent of laser-cooling and trapping of atoms \cite{Raa87,Phi82}, followed shortly by the realization of 
Bose-Einstein condensates \cite{And95,davis1995PRL} and quantum degenerate Fermi gases \cite{Dem99}, ultracold quantum gases 
have emerged as a new class of material systems for studies spanning many of the subdisciplines of the physical sciences 
\cite{Blo12,Bra06,Sal12}. 
In the past few years, this progress has increasingly been translated into promising prospects for controlling atomic 
behavior. Such tools as magnetically tuned scattering resonances \cite{Feshbach98,chin2010rmp,koehler2006rmp}, 
periodic lattices of micrometer-sized optical 
potentials \cite{Lew12}, artificial gauge fields \cite{Lin09}, and methods to constrain the dimensionality of the gases 
\cite{Lew12} offer often unprecedented control of the external and interatomic potentials of the atomic gases. 
The purity and accessibly of the ultracold gases have thus enabled the development of the next generation of atomic clocks with record 
accuracy and stability \cite{RevModPhys.87.637} and state-of-the-art inertial force sensors based on matter-wave interferometers 
\cite{Cro09}.

We believe that a broad new range of opportunities exists by exploring
ultracold molecular gases in space, from both fundamental and technological perspectives \cite{safronova2018rmp}. 
Evidently, molecules are more complicated 
quantum objects, and that is exactly where the new opportunities reside. 
But not all molecules are created equal. The particular class of molecules we believe that can have a significant 
impact on the research opportunities provided by a microgravity environment are the so-called Feshbach molecules \cite{chin2010rmp,koehler2006rmp,PhysRevLett.91.250401}. 
These molecules have large spatial extent and are extremely weakly bound, both aspects uniquely collaborate to their 
susceptibility for quantum control and applications to fundamental science and technology. 
For instance, Feshbach molecules formed in Fermionic gases were crucial for revealing fundamental aspects of the
BEC-BCS crossover physics \cite{greiner2003Nature,cubizolles2003prl,regal2004prla,jochim2003Science,
zwierlien2004prl,strecker2003prl,regal2004prlb,chin2005Science,zwierlein2005Nature} while
their heteronuclear counterparts are today an important ingredient for the creation of ultracold polar molecules 
\cite{ospelkaus2008NatPhys,ni2008Science,zirbel2008prl,wu2012prl,heo2012pra,tung2013pra,repp2013pra,
koppinger2014pra,wang2013pra,takekoshi2012pra,deh2010pra} with potential applications to quantum information \cite{ni2018cs,sawant2020njp} and quantum chemistry \cite {son2021control}.

Without doubt, realizing molecular quantum gases has its own pitfalls. In fact, serious limitations on their stability
and lifetime can compromise their use as a viable tool for quantum sciences. In the past few years, however,
greater attention \cite{national2011Recapturing,Turyshev:2007qy,Bin09} and investigation \cite{Mun13,Zoe10} have been given 
to the possibility of producing ultracold molecular gases in microgravity. 
This focus is driven by the fact that many of the environmental and practical limitations, on Earth, for reaching ever 
lower energy and higher symmetry regimes, and for extended atom-atom and atom-photon coherences for these gases, 
are suppressed or removed altogether in microgravity environments such as NASA's Cold Atom Lab (CAL) facility
operating on the International Space Station (ISS) \cite{CAL_nature_20} or the upcoming ISS-based Bose Einstein Condensate Cold Atom Lab (BECCAL) \cite{BECCAL}. In such spaceborne cold atom facilities, one can expect to achieve quantum degenerate gases at densities and temperatures orders of magnitude lower than comparable gases produced on earth.
This is a game-changer for future explorations of molecular studies since, in such environment, molecules will not only suffer 
much less from the limiting factors, but it also opens new regimes for study of low-energy molecules with exceptionally large spatial extents.
Building on the opportunities summarized in this article, we expect that a research program to mature the study and use of ultracold molecules in microgravity will lead a significant advancement of quantum science and technology for 
the decades to come.

%\textcolor{red}{JPD: should we remove the following sentence or this is our disclaimer? Notably, these concepts were also submitted as a white paper for consideration of the Decadal Survey on Biological and Physical Sciences Research in Space 2023-2032 to inform NASA on the long-term promise we envision for studying ultracold molecules in microgravity.}
%JRW: You are correct Jose, we should remove this statement

\section{Challenges and Microgravity Justification}

Association and dissociation of ultracold Feshbach molecules have been enabling probes of fundamental physics 
throughout the last decade \cite{chin2010rmp,koehler2006rmp}. Produced near Feshbach resonances, these molecules are
magnetically tunable and can have large spatial extents and extremely weak binding energies. 
Under typical conditions found with terrestrial experiments, ultracold molecular gases can be highly 
unstable due to collisions leading to molecular de-excitation as well as thermal fluctuations
that can lead to molecular dissociation. In both cases, the timescale for molecular losses 
leaves only a small fraction of time for the system to develop correlations, without which no useful 
physical measurement and application can be realized. The absence of gravity resolves some of these
issues, as it allows for ultralow densities and ultralow temperatures thus preventing collisional losses.
Further, well-overlapped dual-species gases, necessary for formation of heteronuclear molecules, 
are generally prohibited in weak traps on earth due to ``gravitational sag" \cite{BECCAL}. In microgravity, well overlapped gases in weak traps can be created with extended lifetime allowing 
for a more efficient manipulation of the sample. 

However, at ultralow densities molecular association will tend to be challenging due to the lack of good overlap 
between atom and molecular states. This will require to consider association schemes adapted to such
unique environments. In fact, recent theoretical studies point to promising new ways to achieve high molecular 
association efficiency \cite{dincao2017pra,waiblinger2021pra} in a microgrativity environment.
That, along with novel molecular cooling techniques enabled in space, facilitates high-space densities
for better and more accurate measurements. Furhtermore, the removal of a linear gravitational potential combined with the long free floating times would allow for enhanced 
delta-kick cooling and adiabatic decompression. These two techniques are key to reach extremely low kinetic temperatures at the picoKelvin level, yet conserving the phase-space density of the sample \cite{ammann1997,PRLKovachy2015,PRLMuntinga2013,myrskog2000pra,leanhardt2003sci,Deppner2021b}, 
opening the door to a new parameter regime of ultralow densities and ultracold temperatures.

\section{Opportunities}

From our perspective, there exist a number of exciting opportunities one can reach by utilizing molecular physics as a tool 
to advance both technological and fundamental aspects of ultracold quantum gases. 
We elaborate our perspective below, and discuss the ways in which the access of molecular 
physics can provide strong impact to space-based fundamental 
physics research for the advancement of quantum science and technologies.
%For the reasons elaborated below, we 
%believe that accessing molecular physics should be a main focus of space-based fundamental 
%physics research from 2023-2032 
%for the advancement of quantum science and technologies. 
The molecular bound is the key element in which studies with ultracold molecular gases can take advantage to
probe a number for fundamental properties of matter and develop novel technologies. 
Molecules offer a new path to generate correlation and entanglement between different atomic species but also are
themselves a laboratory to study various fundamental physics aspects.
Here we will list and briefly discuss a number of the directions whether they are of a more fundamental or applied nature.

\paragraph{Dual-species Atom Interferometry for WEP}

As the precision and ultimate accuracy of atom-interferometers (AIs) improve, greater control of the momentum and position 
profiles of the atomic clouds are required to minimize systematic shifts attributed to external fields and forces~\cite{Gaaloul2014,Loriani2020}.
These requirements are compounded for comparisons of dual-species atom interferometers that are gaining interest
for precision tests of Einstein’s General Relativity theory and searching for violations of the Weak Equivalence Principle (WEP)
\cite{Bon13}. Such measurements provide complimentary tests of the universality of free-fall with quantum objects and further 
bound the theories that predict violations of the Equivalence Principle, in the pursuit of incorporating gravity with the standard 
model.

Ideally, WEP experiments with AIs would probe the phase evolution of two distinct species of ultracold gases in the absence of all 
perturbing forces except gravity. Although microgravity environments provide one of the best conditions for such precision 
measurements, at the level of precision achievable with AIs in microgravity the largest source of systematic error for 
precision tests of the WEP could still come from the deviation of the center-of-mass 
trajectories of the two gases in the presence of an unknown gravity gradient \cite{PhysRevD.102.124043,Chiow2017GravitygradientSI}. This is exactly where Feshbach molecules can play a
major role. 
By forming a purely molecular gas and subsequently dissociating Fehbach molecules, one creates a highly correlated 
heteronuclear atomic gas in both their position and momentum distributions which drastically reduces the two clouds center-of-mass 
displacement. 
The utility of this preparation scheme arises from the facts that, before dissociation, the size of the Feshbach molecule 
is controllable, thus allowing atoms of different species to be at the same mean distance. 
At dissociation, the $s$-wave and weakly bound character of the Feshbach molecule provides a final velocity 
distribution that is spatially symmetric and highly correlated between. In this way, much of the shot-to-shot fluctuations 
in the preparation of each species becomes common-mode and cancels in differential atom interferometry measurement. 

We believe that, in conjunction with gravity demodulation \cite{PhysRevD.102.124043,Chiow2017GravitygradientSI} and demonstrated compensation \cite{Roura2014} schemes,  state preparation using Feshbach molecules will mitigate static and dynamic systematics stemming from center of mass offsets of multiple atomic test masses and will open the door for the next generation of 
differential measurements of atom interferometry with dual atomic species. The research is important not only for advancing the 
technology of these space-based precision sensors but also for fundamental advances for future tests of the WEP with unprecedented accuracy. 

\paragraph{Molecular Entanglement for Atom Interferometry}
Controlled dissociation of weakly bound diatomic molecules near a Feshbach resonance provides a highly versatile path for generating atoms in motionally entangled states with a potential 
for high impact. The resulting pair-correlated atoms have uses in quantum information, precision measurements, and fundamental tests of quantum mechanics 
\cite{fry1995pra,poulsen2001pra,kherunssyan2002pra,kheruntsyan2005pra,yurovsky2003pra,
savage2007prl,kheruntsyan2005prl,zhao2007pra,davis2008pra,gneiting2008prl,gneiring2010pra}.
In particular, such entangled states may enable substantial gains in sensitivity of matter wave interference–based 
instruments as well as fundamental studies of quantum phase transitions \cite{orzel2001sci}.

An exciting new perspective is obtained by also considering other forms of molecular states besides Feshbach molecules.
The association and dissociation of homonuclear or heteronuclear Efimov triatomic molecules 
\cite{dincao2018JPB,barontini2009prl,bloom2013prl,tung2014prl,pires2014prl,maier2015prl} and other more exotic atom-molecule 
pairs \cite{lin2020Nature}, for instance, provide a way to produce multiparticle entanglement between massive objects that have a broad-range of applications.
Although highly correlated quantum states can be detected through the measurement of atom shot noise correlations 
\cite{altman2004pra,belzig2005aip,bach2004prl,greiner2005prl},
the capability to perform precision interferometry with atoms in microgravity also opens up the possibility to use interferometry as a tool for precise characterization of the entangled states themselves. 
The long interrogation times for atom-interferometry available in microgravity and the corresponding favorable conditions to associate 
atoms to more complex molecules will allow one to characterize and manipulate properties of entangled few-body states from a much 
deeper and fundamental perspective. More generally, as shown in Ref.~\cite{fletcher2017sci},
interferometric studies can provide ways to characterize the coherent evolution of strongly interacting ultracold 
matter far from equilibrium. 

\paragraph{Ultra-low temperatures for quantum mixtures}
Since Feshbach molecules are lying in the asymptotic region of the atom-atom potential, a far-detuned optical dipole trap (ODT) would only induce molecular transitions between states in this ``almost-atomic" region.
It is therefore possible to approximate the polarizability of a Feshbach molecule by the sum of polarisabilities of the relative atomic counterparts. Doing so, it is possible to define a dipole trap potential of a molecule that could be used to manipulate it as a whole. Techniques of delta-kick collimation that were successful to realize picoKelvin energies for atomic samples~\cite{Deppner2021b} could be transferred here to slow down the free expansion of quantum mixtures. This circumvents the need for multiple pulses to delta-kick binary ensembles of different atomic masses as proposed in~\cite{NJP_Corgier_2020}. When decoupled, these ultra-cold quantum mixtures could be used as an input state in several applications such as a WEP test.  
\paragraph{Molecular Interferometry}
Following the same reasoning as in the previous paragraph, 1D optical lattices could be generated for the molecular state. An example 
implementation is a retro-reflected, red-detuned laser with at-least 2 frequency components that can be independently controlled (frequency 
and amplitude for each tone) and ramped with respect to each other within an experimental run. When this light grating is appropriately accelerated~\cite{Fitzek2020}, Bragg or Bloch diffraction of a Feshbach molecule can occur. The field of matter-wave interferometry with diatomic (composite) molecules could be created with a lot more possibilities than the atomic counterpart thanks to the several additional molecular degrees of freedom and the hight control one has over them. Applications in quantum sensing, fundamental tests and many-body physics are anticipated.

\paragraph{Quantum mechanics tests}
%Molecular states help to generate entanglement therefore could help in testing interferometric and non-interferometric CSL models.

To this day, macroscopic quantum interference experiments are the commonly accepted and proven method to probe the validity of quantum theory at the classical boundary. With every successful interference measurement, one falsifies hypothetical modifications of the Schr\"odinger equation that would predict a spontaneous collapse of quantum states at the probed system scale \cite{bassi2013models,schrinski2019macroscopicity}.

At the moment, the most macroscopic matter-wave platforms are Bragg pulse atom interferometers with seconds of interference time and delocalizations up to the meter scale~\cite{kovachy2015quantum,xu2019probing,schrinski2020quantum} and near-field interferometers with molecules of more than ten thousand atomic mass units~\cite{fein2019quantum}, albeit at smaller time and length scales. Molecule interferometers are generally expected to take the lead in the long run, because their sensitivity to well-studied collapse models such as the Continuous Spontaneous Localization (CSL) model \cite{ghirardi1990markov} typically amplifies with the square of the particle mass and only linearly with interference time.

However, with precisely controllable diatomic interactions at hand, interfering condensates could be brought to an entangled state and thereby achieve the same quadratic scaling of CSL sensitivity with the total condensate mass \cite{schrinski2020rule}. Moreover, the here envisaged loosely bound Feshbach diatoms at ultra-low temperatures would serve as a highly sensitive probe for collapse-induced spontaneous dissociation at unprecedented scales of binding length and energy, complementing other non-interferometric CSL test platforms like the LISA-pathfinder mission \cite{carlesso2016experimental}. Thus Mach-Zehnder-like interferometry with diatomic molecules could complement and potentially surpass existing interferometric and non-interferometric platforms to test quantum theory against collapse models or other fundamental decoherence effects.

\paragraph{Molecules and variation of fundamental constants}

Finally, a space environment enables homonuclear and heteronuclear molecules to be prepared at unprecedented low energies with weak binding 
energies, large mean radii, and strong interactions. Experimental tests of universal theories in this regime will provide a unique window 
into the fundamental nature of our universe. Feshbach molecules allow for tests for variations of fundamental constants with unprecedented sensitivity \cite{chin2006prl,chin2009njp,borschevsky2011pra,gacesa2014jms}. In particular, the precise measurement of properties
of Feshbach molecules is extremely sensitive to the variation of the electron-to-proton mass ratio, thus providing 
a precision test of the grand unification models discussed in Refs. \cite{marciano1984prl,calmet2002epj,langacker2002plb}.
Due to the ultralow density regime allowed in microgravity, Feshbach molecules can be made substantially larger than those
in Earth-bond experiments. This can lead to a major advance of testing the variation of fundamental constants.

\section{Conclusions}
Efforts to mature the technology for space-enabled studies of ultracold molecular physics can be performed simultaneously on various microgravity platforms. Demonstrations using the ZARM droptower in Bremen or the Einstein Elevator in Hannover, DE, can be used for initial studies of the opportunities outlined above in preparation for dedicated flight experiments. NASA's CAL is already being utilized onboard the ISS to optimize association and dissociation of ultracold heteronuclear Feshbach molecules at ultra-low temperatures, aiming for even nanoKelvin scales. Subsequent studies using CAL and BECCAL are also planned. Follow-on missions to BECCAL could then utilize matured technologies of ultracold molecular physics in space for transformative science.

The science opportunities outlined in this manuscript must be done in space to achieve their ultimate performance to address fundamental research questions including:
\begin{itemize}
  %\item What are the quantum properties of atoms and molecules?
  \item How is entanglement influenced by gravity and the intrinsic properties of the quantum system?
  \item How does complexity and order arise from quantum interactions?
  \item Is Einstein’s General Relativity valid under all experimental conditions?
\end{itemize}
We anticipate that as the answers to these fundamental questions become available and the toolbox of quantum technologies from ultracold molecules in space is matured, the transformative nature of the research will also have impact on human exploration and far-reaching value to everyday life for humans on Earth.\\

\acknowledgments

A portion of this research was carried out under a contract with the National Aeronautics and Space Administration. 
%JRW and JPD were supported by JPL-CALTECH under NASA contract NNH13ZTT002N. 
JPD also acknowledges partial support from the U. S. National Science Foundation, grant number PHY-2012125. U.S. Government sponsorship acknowledged.

\bibliography{References}

\end{document}